\documentstyle[preprint,aps,floats]{revtex}    
\input epsf
\begin{document}
\preprint{$
\begin{array}{l}
\mbox{CWRU-P31-99}\\
\mbox{UB-HET-99-02}\\
\mbox{September~1999} \\ [1cm]
\end{array}
$}
\tightenlines
\title { Zero Zeros After All These (20) Years\footnote{\it Talk given by RB 
at the Workshop ``Formfactors from Low to High $Q^2$'' (Brodskyfest),
University of Georgia, Athens, Georgia, 
Sept. 17, 1999, to appear in the Proceedings}}
\author
{ Ulrich Baur}
\address{Department of Physics, State University of New
York at Buffalo, Buffalo, NY 14260-1500} 
\author{Robert Brown}
\address{Department of Physics, Case Western Reserve
University,
Cleveland, OH 44106-7079}
\maketitle
\begin{abstract}
We celebrate two birthdays connected with the radiation zero
phenomenon. First, a striking dip in the theoretical angular distributions of
radiative weak-boson production was discovered twenty years ago. The key
experimental interest is that this will not occur in any deviation from the
standard model. Second, the classical training of Stanley Brodsky began 
sixty years ago, which was instrumental in understanding why theoretical 
spin-independent radiation zeros appear in almost all Born amplitudes
for the radiation of photons and gluons and other massless gauge bosons 
(but rarely in physical kinematic regions). And there are approximate
zeros for massive bosons and ``Type II'' zeros that can also be
studied. We discuss how the difficulties in observing the original 
Mikaelian-Samuel-Sahdev zero finally may be surmounted next year.

\end{abstract}

\section*{Original Zero}

The first radiation zeros were discovered as a consequence of
a general investigation~\cite{BM,BSM}
of the production of electroweak pairs in 
hadronic 
collisions
\begin{equation}
p \overline{p},\; pp \rightarrow WW,ZZ,WZ,W\gamma \;\;+\;\;X
\end{equation}
and neutrino reactions
\begin{equation}
\nu e \rightarrow WZ,W\gamma 
\end{equation}
addressed to very-high-energy cosmic-ray physics.
The investigation probed the sensitivity of these 
reactions to the trilinear gauge boson couplings 
and, at the present time, useful limits on their 
deviations from gauge theory predictions have been obtained 
in $e^+e^-$ and hadron collider 
experiments \cite{wudka}.  Pronounced dips were found~\cite{BSM} 
in the angular 
distributions for the production of $W\gamma$ and $WZ$ in the two-body parton 
and lepton reactions. Subsequent work~\cite{MSS} by Mikaelian, 
Samuel, and Sahdev proved that the $W\gamma$ dips were in fact 
exact zeros at particular angles, which would be ruined by 
non-gauge couplings.  The Wisconsin brain trust~\cite{DGHL} 
followed with a general demonstration of
the implied amplitude factorization. 

\section*{Simple Zero}

Since it is so easy to do so, we show how a radiation zero 
arises in lowest-order radiative charged-scalar fusion
\begin{equation}
{\rm scalar \; 1 + scalar \; 2} \rightarrow {\rm scalar\; 3 + photon}
\end{equation}
whose Feynman diagrams lead to the Born amplitude
\begin{equation}
M_{\gamma}^{sc} =  \frac{Q_3}{p_3\cdot q} p_3 \cdot \epsilon 
  -\frac{Q_1}{p_1\cdot q} p_1 \cdot \epsilon 
   -\frac{Q_2}{p_2\cdot q} p_2 \cdot \epsilon
\end{equation}
with charges $Q_i$, four-momenta $p_i$, and photon momentum $q$ and 
polarization $\epsilon$.  (The trilinear scalar coupling is taken 
to be unity.)  Using momentum conservation, we observe that 
$M_{\gamma}^{sc} = 0$ if all $Q_i/p_i\cdot q$ are the same.  
This is exactly the same condition found for the 
$W\gamma$ amplitude, independent of the spins.  It leads to a zero when
\begin{equation}
\cos \theta^* =  \frac{Q_1-Q_2}{Q_1+Q_2} 
\end{equation}
for the center-of-mass angle $\theta^*$ of particle 3 relative to the
direction of particle~1 (or between the
photon and particle~2).  This reduces to 
$\cos \theta^* = 1/3 \;\;(-1/3)$ for  
$u\overline{d} \rightarrow W^+\gamma$ ($d\overline{u} \rightarrow W^-\gamma$).

\section*{Zeros Everywhere:  Theorems}

Faced with the vanishing of the above 
amplitudes, Brodsky asserted that 
the way to look at these zeros was as the complete destructive 
interference of classical radiation patterns.  
Following this lead, and making a long story 
short (see the longer story with better 
referencing in~\cite{B}), an arbitrary 
number $n$ of external charged particles 
was considered and a general set 
of radiation interference theorems were found~\cite{BKB}. Again 
considering the emission 
of a photon with momentum $q$, the tree amplitude approximation 
vanishes, independent
of any particle's spin $\leq 1$, for common
charge-to-light-cone-energy
ratios (the radiation null zone),
\begin{eqnarray}
M_{\gamma}(tree) & = & 0\nonumber\\
{\rm if}\; \frac{Q_i}{p_i \cdot q} & = & {\rm same,\; all}\; i
\end{eqnarray}
where the $i^{th}$ particle has electric charge $Q_i$ and four-momentum
$p_i$.  All couplings must be prescribed by local gauge theory.
We see why it took so long to discover radiation zeros since 
the first null zone requirement is that all charges must 
have the same sign.   Fractionally
charged quarks and weak bosons are needed in order to 
get three things:  Same-sign charges, a process
well-approximated by a Born amplitude, and a four-particle 
reaction so the null zone was simple. While there are zeros 
associated with any gauge group when the corresponding 
massless gauge bosons are emitted, in QCD, color charges
are averaged or summed over in hadronic reactions.  In thinking of the weak
bosons themselves, electroweak symmetry is broken and nonzero weak-boson masses ruin radiation interference.   

What about other photonic zeros?  The zero in electron-electron
bremsstrahlung is less interesting.  Zeros in electron-quark and
quark-antiquark bremsstrahlung require jet identification 
along with the more complicated phase space~\cite{BIL}.
While we shall say more about other tests, we find 
ourselves returning again and again to the original $W\gamma$ zero. 

\section*{Zero $\neq$ Zero}

There are various corrections that turn the $W\gamma$ zero into a dip.  
Theoretically, higher-order (closed-loop) corrections
will not vanish in the null zone, since the internal loop momenta
cannot be fixed.  Structure function effects, higher order QCD
corrections, finite $W$ width effects, and photon radiation from the
final state lepton line all tend to fill in the dip.

The main complication in the extraction of the $\cos\theta^*$
distribution in $W\gamma$ production, however, originates from the
finite resolution of the detector and
ambiguities in reconstructing the parton center of mass frame. The
ambiguities are associated with the
nonobservation of the neutrino arising from $W$ decay. Identifying the
missing transverse momentum, $p\llap/_T$, with the transverse momentum of the
neutrino of a given $W\gamma$ event, the unobservable longitudinal neutrino
momentum, $p_L(\nu)$, and thus the parton center of
mass frame, can be reconstructed by imposing the constraint that the
neutrino and charged lepton four momenta combine to form the $W$ rest
mass. The resulting quadratic equation, in general, has
two solutions. In the approximation of a zero $W$-decay width, one of
the two solutions coincides with the true $p_L(\nu)$. On an event by
event basis, however, it is impossible to tell which of the two
solutions is the correct one. This ambiguity
considerably smears out the dip caused by the amplitude zero. Problems
associated with the reconstruction of the parton center of mass frame
could be avoided by considering hadronic $W$ decays. The horrendous
QCD background, however, renders this channel useless.

\section*{Zero Progress}

\begin{figure} 
\begin{center}
\begin{tabular}{c}
\\[-4.5cm]
\epsfysize=10.cm
\epsffile{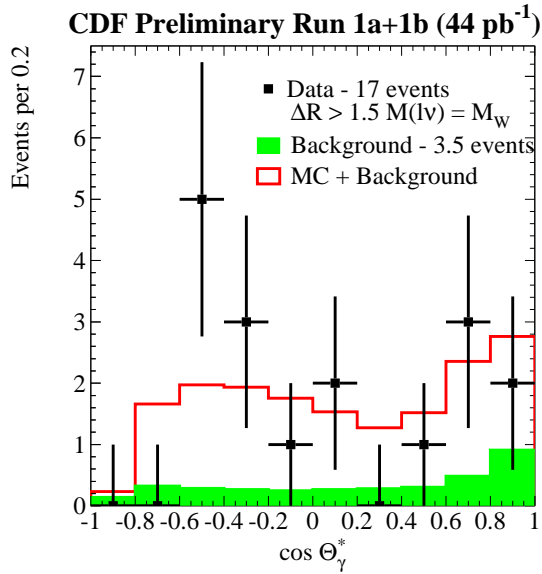}
\end{tabular}
\\[1.6cm]
\caption{Preliminary CDF results for the $\cos\theta^*$ distribution
obtained from a partial data set of the 1992-95 Tevatron run 
(from Ref.~[9]). 
The points are the data and the open histogram is the sum of
the SM prediction and the background.  The shaded histogram is the
background estimate.} 
\label{FIG:FOURTEEN2}
\end{center}
\end{figure}

At present there is only a preliminary study by the CDF collaboration of
the $\cos\theta^*$ distribution~\cite{Ben} from a partial data sample of 
the 1992-95 Tevatron run. 
The event rate is still insufficient to make a
statistically significant statement about the existence of the radiation
zero.  One can at best say that ``there is a hint of gauge zero'' in 
Fig.~\ref{FIG:FOURTEEN2}.


\section*{Zero Help}

\begin{figure}
\begin{center}
\begin{tabular}{c}
\\[-2.cm]
\epsfysize=13.cm
\epsffile{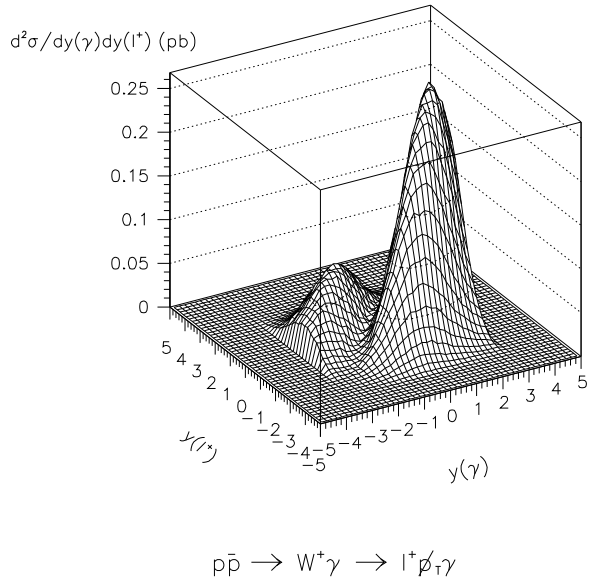}
\end{tabular}\\[-3.cm]
\caption{The double differential distribution $d^2\sigma/dy(\gamma)
dy(\ell)$ for $p\bar p\to W^+\gamma\to\ell p\llap/_T\gamma$ at the
Tevatron.}
\label{FIG:FOURTEEN}
\end{center}
\end{figure}

Instead of trying to reconstruct the parton center of mass frame and
measure the $\cos\theta^*$ or the equivalent rapidity distribution in
the center of mass frame, one
can study rapidity correlations between the observable final state
particles in the laboratory frame. Knowledge of the neutrino
longitudinal momentum is not required in determining this
distribution.
Event mis-reconstruction problems originating from the two possible
solutions for $p_L(\nu)$ are thus automatically avoided.
In $2\to 2$ reactions differences of rapidities are invariant
under boosts. One therefore expects that the double differential
distribution of the rapidities, $d^2\sigma/dy(\gamma)dy(W)$, where
$y(W)$ and $y(\gamma)$ are the $W$ and photon rapidity, respectively,
in the
laboratory frame, exhibits a `valley,' signaling the SM amplitude
zero~\cite{BEL}. In
$W^\pm\gamma$ production, the dominant $W$ helicity is $\lambda_W =
\pm 1$~\cite{BBS}, implying that
the charged lepton, $\ell=e,\,\mu$, from $W\to\ell\nu$ tends to be
emitted
in the direction of the parent
$W$, and thus reflects most of its kinematic properties. As a result,
the valley signaling the SM radiation zero should manifest itself also
in the $d^2\sigma/dy(\gamma)dy(\ell)$ distribution of the photon and
lepton rapidities. The theoretical prediction of the
$d^2\sigma/dy(\gamma)dy(\ell)$ distribution in the Born approximation
for $p\bar p$ collisions at 1.8~TeV is
shown in Fig.~\ref{FIG:FOURTEEN} and indeed exhibits a pronounced
valley
for rapidities satisfying $\Delta y(\gamma,\ell)=y(\gamma)-y(\ell)
\approx -0.3$. The location of the valley can be easily understood
from
the value of $\cos\theta^*$ for which the zero occurs and the average
difference between the $W$ rapidity and the rapidity of the $W$ decay
lepton~\cite{BEL}.

To simulate detector response, transverse momentum cuts of
$p_T(\gamma)
>5$~GeV, $p_T(\ell)>20$~GeV and $p\llap/_T>20$~GeV, rapidity cuts of
$|y(\gamma)|<3$ and $|y(\ell)|<3.5$, a cluster transverse mass cut
of
$m_T(\ell\gamma;p\llap/_T)>90$~GeV and a lepton-photon separation cut
of $\Delta R(\gamma,\ell)>0.7$ have been imposed in the Figure. Here,
$\Delta R(\gamma,\ell)$ is the separation between the lepton and the
photon in the azimuthal angle-pseudorapidity plane, 
\begin{equation}
\Delta
R(\gamma,\ell)=\sqrt{\Delta\Phi(\gamma,\ell)^2+\Delta\eta(\gamma,\ell)^2}.
\end{equation}
The cluster
transverse mass cut suppresses final state photon radiation which
tends
to obscure the dip caused by the radiation zero. For 10~fb$^{-1}$, a
sufficient number of events should be available to map out $d^2\sigma/
dy(\gamma)dy(\ell)$ in future Tevatron experiments.

\begin{figure}
\begin{center}
\begin{tabular}{c}
\epsfysize=8.5cm
\epsffile{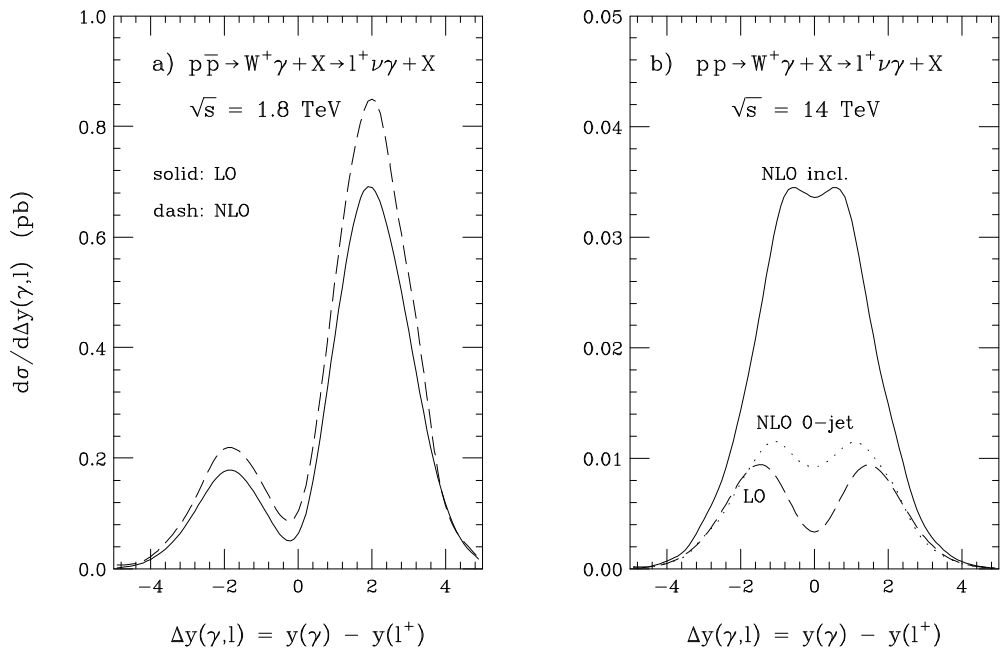}
\end{tabular}
\caption{Photon lepton rapidity difference distribution for $W\gamma$
production in the SM at a) the Tevatron and b) the LHC.}
\label{FIG:NINE}
\end{center}
\end{figure}

For smaller data sets, the rapidity
difference distribution, $d\sigma/d\Delta y(\gamma,\ell)$, is a more
useful variable. In the photon lepton rapidity difference
distribution, the SM radiation zero leads to a strong dip located
at $\Delta y(\gamma,\ell)\approx -0.3$~\cite{BEL}.
The LO and NLO predictions of the SM $\Delta y(\gamma,\ell)$
differential cross section for $p\bar p\to\ell^+p\llap/_T\gamma$ at
the Tevatron are shown in Fig.~\ref{FIG:NINE}a.
Next-to-leading QCD corrections leave a reasonably visible dip. 

In $pp$ collisions, the dip signaling the amplitude zero is shifted to
$\Delta y(\gamma,\ell)=0$. Because of the increased $qg$ luminosity, the
inclusive QCD corrections are very large for $W\gamma$
production at multi-TeV hadron colliders~\cite{NLOWGAM}. At the
LHC, they enhance the cross section by a factor 2~--~3. The rapidity
difference distribution for $W^+\gamma$ production in the SM
for $pp$ collisions at $\sqrt{s}=14$~TeV is shown in
Fig.~\ref{FIG:NINE}b. Here we have
imposed the following lepton and photon detection cuts:
\begin{eqnarray}
p_T(\gamma) > 100~{\rm GeV/c}, & \qquad & |\eta(\gamma)|<2.5, \\
p_T(\ell)>25~{\rm GeV/c}, &\qquad &|\eta(\ell)|<3, \\
p\llap/_T>50~{\rm GeV/c}, &\qquad &\Delta R(\gamma, \ell)>0.7.
\end{eqnarray}
The inclusive NLO QCD corrections are seen to considerably obscure
the amplitude zero. The bulk of the corrections at LHC energies
originates from quark-gluon fusion and the kinematical region where
the photon is produced at large $p_T$ and recoils against a
quark, which radiates a soft $W$ boson which is almost collinear to
the quark. Events which originate from this phase space region usually
contain a high $p_T$ jet. A jet veto therefore helps to reduce the QCD
corrections. Nevertheless, the remaining QCD corrections
still substantially blur the visibility of the radiation zero in
$W\gamma$ production at the LHC~\cite{BEL}.

Given a sufficiently large integrated luminosity, experiments at the
Tevatron studying lepton-photon rapidity correlations therefore offer
a {\sl unique} chance to observe the SM radiation zero in $W\gamma$
production.
Nonstandard $WW\gamma$ couplings tend to fill in the dip in the
$\Delta
y(\gamma,\ell)$ distribution caused by the radiation zero.

Indirectly, the radiation zero can also be observed in the $Z\gamma$
to $W\gamma$ cross section ratio~\cite{BEO}. Many theoretical and
experimental uncertainties at least partially cancel in the cross
section
ratio. On the other hand, in searching for the effects of the SM
radiation zero in the $Z\gamma$ to $W\gamma$ cross section ratio, one
has to assume that the SM is valid for $Z\gamma$ production. Since the
radiation zero occurs at a large scattering angle, the photon $E_T$
distribution in $W\gamma$ production falls much more rapidly than that
of photons in $Z\gamma$ production. As a result, the SM $W\gamma$ to
$Z\gamma$ event ratio as a
function of the photon transverse energy, $E^\gamma_T$, drops rapidly.

\section*{Multizeros}
 
\begin{figure}[t]
\begin{center}
\begin{tabular}{c}
\epsfxsize=6.0in\hspace{0in}\epsffile[38 216 553 530]{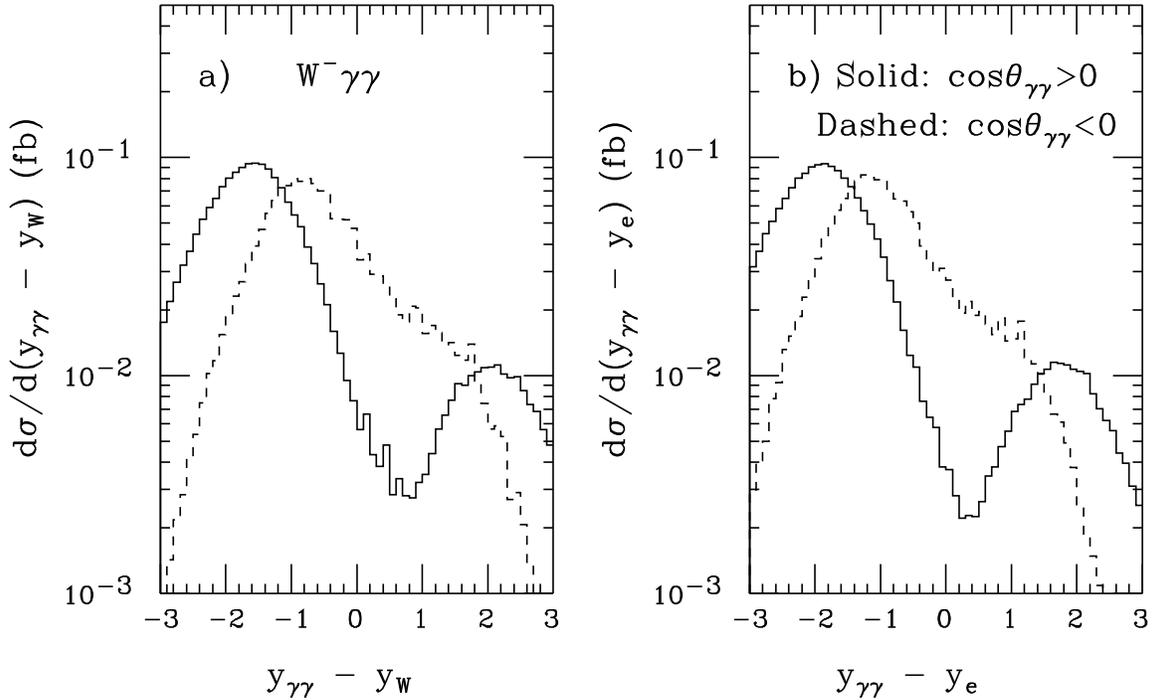}
\end{tabular}
\caption{Rapidity difference distributions for $p\bar p\to e^-\bar
\nu\gamma\gamma$ at $\protect{\sqrt{s}=2}$~TeV. Part (a) shows the
$y_{\gamma\gamma}-y_W$ spectrum, while part (b) displays the
$y_{\gamma\gamma}-y_e$ distribution. The solid (dashed) curves are for
$\cos\theta_{\gamma\gamma}>0$ ($\cos\theta_{\gamma\gamma}<0$). }
\label{FIG:XRAZ0}
\end{center}
\end{figure}

Adding more external photons to a reaction with a Born-amplitude 
radiation zero will still leave us with a null zone which demands,
however, that all photons be collinear~\cite{BKB,BCS}.
In view of the fact that the quadrilinear coupling $WW\gamma\gamma$
contributes, it is of interest to consider the radiation zero in 
$W\gamma\gamma$ production.  
The $\Delta y(\gamma\gamma,W)=y_{\gamma\gamma}-y_W$
distribution is expected to display a clear dip for photons with a 
small opening
angle, $\theta_{\gamma\gamma}$, in the
laboratory frame, {\it i.e.} at $\cos\theta_{\gamma\gamma}\approx 1$.
Calculations show~\cite{BHKSZ} that requiring $\cos\theta_{\gamma\gamma}>0$ 
is already
sufficient. Figure~\ref{FIG:XRAZ0}a displays a pronounced dip in
$d\sigma/d\Delta y(\gamma\gamma,W)$ for $\cos\theta_{\gamma\gamma}>0$
at the Tevatron, located at $\Delta y(\gamma\gamma,W)\approx 0.7$ (solid
line) for $e^-\bar\nu\gamma\gamma$ production at the Tevatron.
In contrast, for $\cos\theta_{\gamma\gamma}<0$, the $\Delta
y(\gamma\gamma,W)$ distribution does not exhibit a dip (dashed line).

In the dip
region, the differential cross section for
$\cos\theta_{\gamma\gamma}<0$
is about one order of magnitude
larger than for $\cos\theta_{\gamma\gamma}>0$. In addition, the
$\Delta y(\gamma\gamma,W)$ distribution extends to significantly
higher
$y_{\gamma\gamma}-y_W$ values if one requires
$\cos\theta_{\gamma\gamma}>0$. This reflects the narrower rapidity
distribution of the two-photon system for
$\cos\theta_{\gamma\gamma}<0$,
due to the larger invariant mass of the system when the two photons are
well separated.

Exactly as in the $W\gamma$ case,
the dominant helicity of the $W$ boson in $W^\pm\gamma\gamma$
production
is $\lambda_W=\pm 1$. One therefore
expects that the distribution of the rapidity difference of the
$\gamma\gamma$ system and the charged lepton is very similar to the
$y_{\gamma\gamma}-y_W$ distribution and would show a clear signal of
the radiation zero for positive values of $\cos\theta_{\gamma\gamma}$.
The $y_{\gamma\gamma}-y_e$ distribution, shown in Fig.~\ref{FIG:XRAZ0}b,
indeed clearly displays these features. Due to the finite difference
between the electron and the $W$ rapidities, the location of the
minimum is slightly shifted.

To simulate the finite acceptance of detectors, we have imposed the
following cuts in Fig.~\ref{FIG:XRAZ0}:
\begin{eqnarray}
\label{EQ:CUTS}\nonumber
p_T(\gamma) &>& 10\ {\rm GeV}, \quad |y_\gamma| <2.5, \quad
\Delta R(\gamma\gamma) > 0.3 \quad {\rm for \ photons}, \\
p_T(e) &>& 15 \ {\rm GeV}, \quad |y_e| <2.5, \quad
\Delta R(e\gamma) > 0.7 \quad {\rm for \ charged \ leptons},
\end{eqnarray}
and
\begin{equation}\label{EQ:PTMISS}
p\llap/_T>15~{\rm GeV}.
\end{equation}
In addition, to suppress the contributions from final photon
radiation,
we have required that
\begin{equation}
M_T(e^-\nu) > 70~{\rm GeV.}
\label{EQ:TRM}
\end{equation}

The characteristic differences between the $\Delta y(\gamma\gamma,e)=
y_{\gamma\gamma}-y_e$ distribution for $\cos\theta_{\gamma\gamma}>0$
and
$\cos\theta_{\gamma\gamma}<0$ are also reflected in the cross section
ratio
\begin{equation}
{\cal R}= {\int_{\Delta y(\gamma\gamma,e)>-1}d\sigma\over
\int_{\Delta y(\gamma\gamma,e)<-1}d\sigma}~,
\end{equation}
which may be useful for small event samples. Many experimental
uncertainties cancel in ${\cal R}$. For $\cos\theta_{\gamma\gamma}>0$
one finds ${\cal R}\approx 0.25$, whereas for
$\cos\theta_{\gamma\gamma}<0$ ${\cal R}\approx 1.06$.

Although we have
restricted the discussion above to $e\nu\gamma\gamma$ production, our
results also apply to $p\bar p\to\mu\nu\gamma\gamma$. NLO QCD
corrections are not expected to obscure the dip signaling the
radiation
zero at the Tevatron, but may significantly reduce its
observability at the LHC. Given a sufficiently large integrated
luminosity, experiments at the Tevatron studying correlations between
the rapidity of the photon pair and the charged lepton therefore offer
an
excellent opportunity to search for the SM radiation zero in hadronic
$W\gamma\gamma$ production.
Unfortunately, for the cuts listed above, the $W\gamma\gamma$ production 
cross section at the Tevatron is only about 2~fb. Thus, in order
to observe the radiation zero in $W\gamma\gamma$ production, an integrated
luminosity of at least $20 - 30$~fb$^{-1}$ is needed.

\section*{Approximately Zero}

At energies much larger than the $Z$ boson mass, one naively expects
that the $Z$ boson in the process
\begin{eqnarray}
q_1 \> \bar q_2 \rightarrow W^\pm \> Z 
\label{EQ:REACT}
\end{eqnarray}
behaves essentially like a photon with unusual couplings to the
fermions. One therefore might suspect that an approximate radiation zero 
is present in $WZ$ production. In Ref.~\cite{BHO} it was demonstrated
that the process $q_1\bar q_2\to W^\pm Z$ indeed exhibits 
an approximate zero
located at
\begin{equation}
\cos\Theta^*\approx\pm {1\over 3}\tan^2\theta_W\approx\pm 0.1,
\end{equation}
where $\Theta^*$ is the scattering angle of the $Z$ boson relative to
the quark direction in the $WZ$ center of mass frame. The approximate
zero is the combined result of
an exact zero in the dominant helicity amplitudes ${\cal M}(\pm,\mp)$,
and
strong gauge cancellations in the remaining amplitudes. At high
energies, only the $(\pm,\mp)$ and $(0,0)$ amplitudes remain nonzero:
\begin{eqnarray}
{\cal M}(\pm,\mp) &\longrightarrow&  {F \over \sin\theta^* }\,
(\lambda_{\rm w} - \cos\theta^*)\,
\Bigl[  (g^{q_1}_{-} - g^{q_2}_{-} ) \cos\theta^*
      - (g^{q_1}_{-} + g^{q_2}_{-} ) \Bigr] \>, \\
\noalign{\vskip 10pt}
{\cal M}(0,0) &\longrightarrow&  {F\over 2} \, \sin\theta^* \,
{M_Z \over M_W} \, (g^{q_2}_{-} -  g^{q_1}_{-}) \>.
\end{eqnarray}
Here, $\lambda_{\rm w}$
denotes the $W$ boson polarization ($\lambda = \pm 1, 0$ for
transverse
and longitudinal polarizations, respectively), and
\begin{eqnarray}
F = C \, {e^2 \over \sqrt{2} \sin \theta_{\rm w} } \>,
\end{eqnarray}
where $C = \delta_{i_1 i_2} \, V_{q_1 q_2}$ and
$\theta^*=\pi-\Theta^*$.
$i_1$ ($i_2$) is the color index of the incoming
quark (antiquark) and $V_{q_1 q_2}$ is the quark mixing matrix
element.
\begin{equation}
g_{-}^f = {T_3^f - Q_f \sin^2\theta_{\rm w}\over
\sin\theta_{\rm w} \cos\theta_{\rm w}}
\end{equation}
is the coupling of the $Z$-boson to left-handed fermions with $T_3^f
= \pm {1 \over 2}$ representing the third component of the weak isospin.
$Q_f$ is the electric charge of the fermion $f$. 

The existence of the zero in ${\cal M}(\pm,\mp)$ at
$\cos\Theta^*\approx\pm 0.1$ is
a direct consequence of the contributing $u$- and $t$-channel fermion
exchange diagrams and the left-handed coupling of the $W$ boson to
fermions.
Unlike the $W^\pm \gamma$ case with its massless photon kinematics,
the
zero has an energy dependence which is, however, rather weak for
energies sufficiently above the $WZ$ mass threshold.

\begin{figure}[t]
\begin{center}
\begin{tabular}{c}
\epsfysize=8.5cm
\epsffile{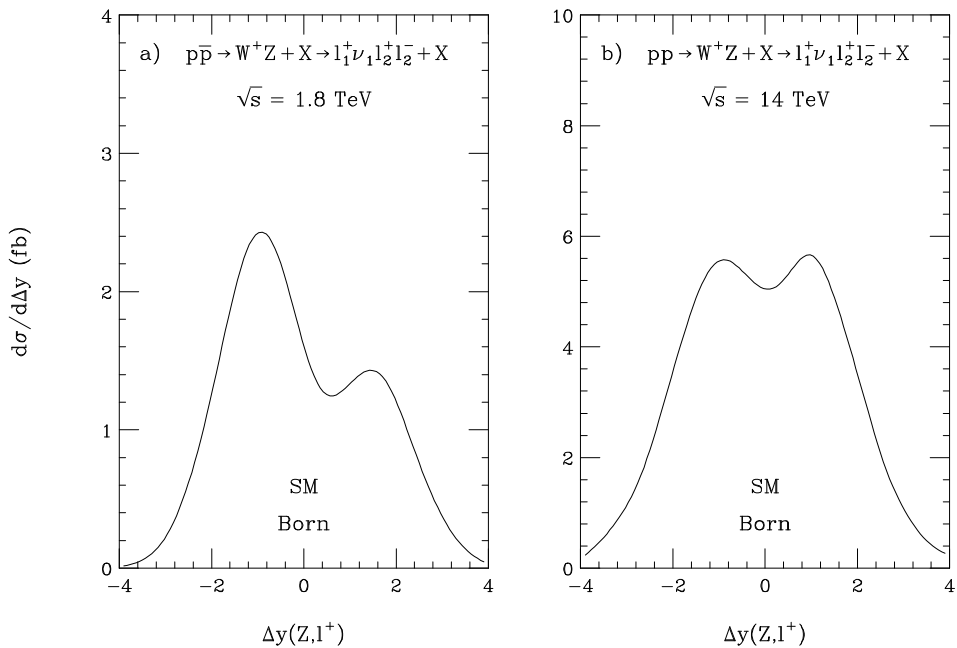}
\end{tabular}
\caption{$Z$ lepton rapidity difference distribution for $WZ$
production in the SM at a) the Tevatron and b) the LHC.}
\label{FIG:RAPDIFF}
\end{center}
\end{figure}

Analogously to the radiation zero in $q_1\bar q_2\to W\gamma$, one can
search for the approximate zero in $WZ$ production in the rapidity
difference distribution $d\sigma/d\Delta y(Z,\ell_1)$~\cite{BHO2},
where
\begin{equation}
\Delta y(Z,\ell_1)=y(Z)-y(\ell_1)
\end{equation}
is the difference between the rapidity of the $Z$ boson, $y(Z)$ and
the
rapidity of the lepton, $\ell_1$ originating from the decay of the $W$
boson, $W\to\ell_1\nu$. The $y(Z)-y(\ell_1)$ distribution for $W^+Z$
production in
the Born approximation is shown in Fig.~\ref{FIG:RAPDIFF}. The
approximate zero in the $WZ$ amplitude leads to a dip in the
$y(Z)-y(W)$
distribution, which is located at $y(Z)-y(W)\approx\pm 0.12$ ($=0$)
for
$W^\pm Z$ production in $p\bar p$ ($pp$) collisions. However, in
contrast
to $W\gamma$ production, none of the $W$ helicities dominates
in $WZ$ production\cite{BBS}. The charged lepton, $\ell_1$, thus only
partly reflects the kinematical
properties of the parent $W$ boson. As a result, a significant part of
the correlation present in the $y(Z)-y(W)$ spectrum is lost, and only
a slight dip survives in the SM $y(Z)-y(\ell_1)$ distribution. This,
and
the much smaller number of $WZ\to\ell_1\nu\ell_2^+\ell_2^-$ events, make
the approximate radiation zero in $WZ$ production much more difficult
to
find at the Tevatron or LHC than the radiation amplitude zero in
$W\gamma$ production.

Due to the nonzero average rapidity difference between the lepton
$\ell_1$ and the parent $W$ boson, the location of
the minimum of the $y(Z)-y(\ell_1)$ distribution in $p\bar p$
collisions is slightly shifted to $y(Z)-y(\ell_1)\approx 0.5$.
In Fig.~\ref{FIG:RAPDIFF}a a rapidity cut of $|\eta(\ell)|< 2.5$ has
been imposed, instead of the cut used in Fig.~\ref{FIG:NINE}a. All
other
rapidity and transverse momentum cuts are as described
before. Furthermore, $\Delta R(\ell,\ell)> 0.4$ is required for
leptons of equal electric charge in Fig.~\ref{FIG:RAPDIFF}.
The significance of the dip in the $y(Z)-y(\ell_1)$ distribution
depends
to some extent on the cut imposed on $p_T(\ell_1)$ and the missing
transverse momentum. Increasing (decreasing) the cut on $p_T(\ell_1)$
($p\llap/_T$) tends to increase the probability that $\ell_1$
is emitted in the flight direction of the
$W$ boson, and thus enhances the significance of the dip. If the
$p\llap/_T>50$~GeV cut at the LHC could be reduced to 20~GeV, the dip
signaling the approximate zero in the $WZ$ production amplitude would
be
strengthened considerably. In contrast to the situation encountered in
$W\gamma$ production, nonstandard $WWZ$ couplings do not always tend
to
fill in the dip caused by the approximate radiation zero. This is due
to the relatively strong interference between standard
and anomalous contributions to the helicity amplitudes for certain
anomalous couplings. As a result, the dip may even become more
pronounced in some cases.

Before we turn to the prospects of observing radiation zeros in the near 
future,
we would like to mention a new development in the general question of 
radiation
zeros. Different kinds of null zones have been found (``Type~II'' radiation
zeros~\cite{S}) in the important process $q\overline{q} 
\rightarrow W^+W^-\gamma$, for which
there are no regular (also called Type~I) zeros. The Type II zeros
require soft photons and
certain coplanarity, but dips survive for harder photons that are
sensitive to the trilinear and quadrilinear gauge boson couplings. 
It will be interesting to see how visible they are in an analysis
incorporating acceptance cuts, detector resolution effects, finite 
$W$ width effects, and decay-lepton radiation. 

\section*{Nonzero Zeros in Zero Zero?}

\begin{figure} 
\begin{center} 
\begin{tabular}{c}
\\[-1.5cm]
\epsfysize=18.cm
\epsffile{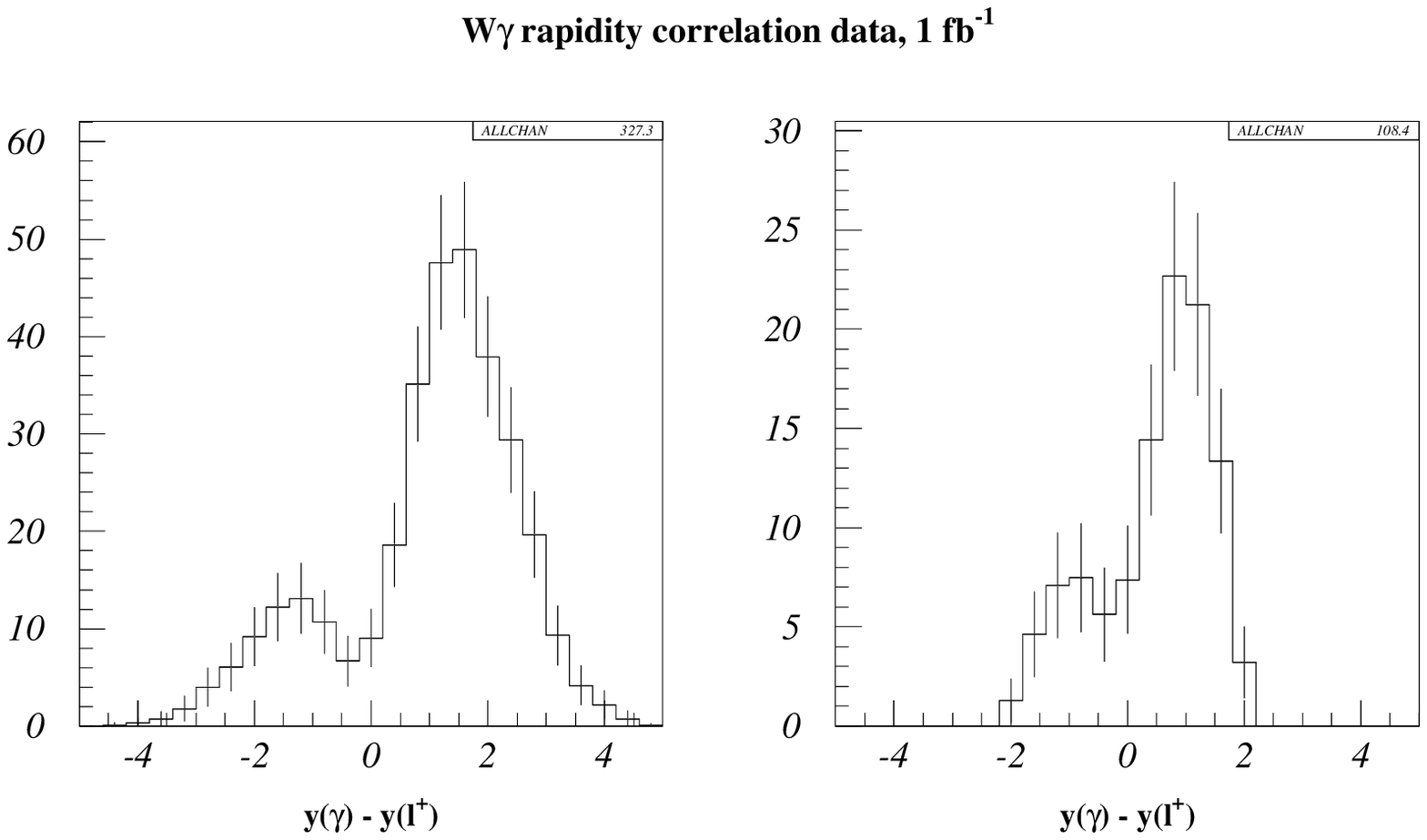}
\end{tabular}\\[-9.3cm]
\centerline{\large a)\hskip 7.cm b)}
\caption{Simulation of the photon lepton rapidity difference 
distribution for $W\gamma$ production at the Tevatron for 1~fb$^{-1}$, a) for
central and endcap photons and electrons, b) for central electrons and
photons only.}
\label{FIG:SIXTEEN}
\end{center}
\end{figure}

How long will we wait for a real dip to appear?
A sufficient
rapidity coverage is essential to observe the radiation zero in 
$d^2\sigma/dy(\gamma)dy(\ell)$ and/or the
$\Delta y(\gamma,\ell)$ distribution~\cite{BEL}. This is
demonstrated in
Fig.~\ref{FIG:SIXTEEN}, which displays simulations of the rapidity
difference distribution for 1~fb$^{-1}$ in the electron channel at the
Tevatron. If both central ($|y|<1.1$) and endcap ($1.5<|y|<2.5$)
electrons and
photons can be used (Fig.~\ref{FIG:SIXTEEN}a), the simulations
indicate
that with integrated luminosities $\geq 1$~fb$^{-1}$ it will be
possible to
conclusively establish the dip in the photon lepton rapidity
difference
distribution which signals the presence of the radiation zero in
$W\gamma$ production. On the other hand, for central electrons and
photons only, the dip is statistically not significant for
1~fb$^{-1}$.
With the detector upgrades which are currently being implemented
for the next Tevatron run, both CDF and D\O\ experiments should have the
capability to analyze the $\Delta y(\gamma,\ell)$ distribution over
the
full rapidity range of $|y|<2.5$.  While the data analysis may take
rather longer, we may hazard the guess that the year Y2K will have more
than three zeros in it.
 
\section*{Zeroing in on Brodsky}

The radiation zeros are the generalization of the well-known
vanishing of classical non-relativistic electric and magnetic dipole
radiation occurring for equal charge/mass ratios (indeed, the low-energy
limit of the null zone conditions) and equal gyro-magnetic 
g-factors.  The null zone is exactly the same as that for 
the completely destructive interference of
radiation by charge lines (a classical
convection current calculation\cite{BKB}) and is preserved
by the fully relativistic quantum Born approximation for gauge theories.

Stan Brodsky has long emphasized the magic of the gyro-magnetic ratio
value $g=2$ predicted by gauge theory for spinor and vector particles.
Only for this value will Born amplitudes have the same null zone as
the classical radiation patterns for soft photons.
And only for this value will 
Born amplitudes have good high-energy behavior. 
In this way we have a connection between the large and small distance
scales, with the value $g=2$ as a bridge.

\section*{Acknowledgment}

It is great to have this chance to acknowledge Stan Brodsky's unique
leadership and collaboration in these radiation matters.  UB would like
to thank S. Errede, T. Han, N. Kauer, G. Landsberg, J. Ohnemus, R.
Sobey and D. Zeppenfeld for pleasant collaboration and
many fruitful discussions.  RB is grateful to K. Kowalski, Sh.
Shvartsman and C. Taylor for advice and collaboration through the
years.  This research was supported in part by NSF grant PHY-9600770.

\end{document}